\documentclass[pra,aps,twocolumn,10pt,showpacs,groupedaddress,superscriptaddress,floatfix]{revtex4-1}
\usepackage{amsmath,amsfonts,amssymb,graphics,graphicx,epsfig,color,times}
\usepackage[utf8x]{inputenc}
\usepackage{color}
\usepackage{bbm, dsfont} 
\usepackage{subfigure}
\usepackage{hyperref}
\usepackage{mathrsfs}
\usepackage{verbatim}
\begin{document}
\title{Cascaded cold atomic ensembles in a diamond configuration as a spectrally entangled multiphoton source}

\author{H. H. Jen}
\email{sappyjen@gmail.com}
\affiliation{Institute of Physics, Academia Sinica, Taipei 11529, Taiwan}
\date{\today}
\renewcommand{\k}{\mathbf{k}}
\renewcommand{\r}{\mathbf{r}}
\newcommand{\f}{\mathbf{f}}
\def\bea{\begin{eqnarray}}
\def\eea{\end{eqnarray}}

\begin{abstract}
We theoretically investigate the spectral entanglement of a multiphoton source generated from the cascade emissions in the cascaded cold atomic ensembles.\ This photon source is highly directional, guaranteed under the four-wave mixing condition, and is also highly frequency-correlated due to finite driving pulse durations and superradiant decay constants.\ We utilize Schmidt decomposition to study the bipartite entanglement of the biphoton states projected from the multiphoton ones.\ This entropy of entanglement can be manipulated by controlling the driving parameters and superradiant decay rates.\ Moreover the projected biphoton states in the cascaded scheme can have larger entanglement than the one produced from only one atomic ensemble, which results from larger capacity in multipartite entanglement.\ This cascaded scheme enables a multiphoton source useful in quantum information processing.\ It also allows for potential applications in multimode quantum communication and spectral shaping of high-dimensional continuous
frequency entanglement. 
\end{abstract}
\maketitle
\section{Introduction}

Quantum computation and quantum information processing \cite{Nielsen2000, Bouwmeester2000} promise to outperform classical implementations for efficient algorithm, secure communication \cite{Gisin2007}, and genuine teleportation of quantum states \cite{Pirandola2015}.\ This superiority of quantumness even envisions a quantum network or quantum internet \cite{Kimble2008} which links various quantum systems to process tasks that are intractable in classical regimes.\ These quantum systems include, to name a few, photonic qubits, trapped ions, atomic ensembles, and solid state systems \cite{Zoller2005,Pirandola2015}, in which unfortunately both strengths and weaknesses coexist \cite{Zoller2005}.\ The fact that there is no prefect quantum system (high efficiency, long coherence time, strong coupling, and scalability, for example) demands an integrability of the well-controlled interfaces between these quantum systems.\ 

A good quantum interface \cite{QIP} involves an efficient generation, distribution, and storage of quantum information.\ One manifestation of these functionalities is long-distance quantum communication \cite{Duan2001} using cold atomic ensembles for a quantum repeater \cite{Briegel1998, Dur1999}.\ The building blocks for this quantum repeater protocol therefore rely on a generation of light-matter entanglement \cite{Matsukevich2004,Chou2004} and storage of it \cite{Chaneliere2005, Chen2006, Laurat2006, Zhao2009} using $\Lambda$-type atomic configurations, making the atomic ensemble (mostly alkali metals) a good candidate for quantum network.\ However one drawback for the fiber-based quantum information transmission is the attenuation loss for $D$ line transitions in alkali metals.\ To reach a minimal-loss optical fiber transmission, telecommunication (telecom) bandwidth from diamond-type atomic configurations \cite{Chaneliere2006, Radnaev2010, Jen2010} serves the purpose and allows for an optimal operation in long-distance quantum communication.\ 

In addition to the advantage of the telecom bandwidth in fiber transmission, continuous frequency entanglement of cascade emissions from diamond-type atomic transitions can be generated \cite{Jen2012-2} and spectrally shaped \cite{Jen2016a, Jen2016b} to create a highly entangled biphoton source.\ This high communication capacity in continuous variables \cite{Braunstein2005} is also present in various degrees of freedom, for example, the transverse momentum \cite{Law2004, Moreau2014}, space \cite{Grad2012}, time \cite{Branning1999, Law2000, Parker2000}, and orbital angular momenta of light \cite{Arnaut2000, Mair2001, Molina2007, Dada2011, Fickler2012, Nicolas2014, Ding2015, Zhou2015}.\ In the perspective of spectral shaping, in Ref. \cite{Jen2016a}, we have investigated the frequency-entangled two-photon state in the scheme of multiplexed cold atomic ensembles.\ This two-photon state is generated from the cascade emissions in the diamond-type atomic configuration.\ Its upper and lower transitions (also denoted as signal and idler) are respectively in the telecom and infrared bandwidths.\ In the multiplexed scheme, a multiple common excitation pulses are applied simultaneously to the respective atomic ensembles along with individually controlled frequency shifters for the cascade emissions.\ Under the condition of weak excitations as in the quantum repeater protocol \cite{Duan2001}, we effectively create a biphoton state with additive spectral functions where individual frequency shifts for the signal and idler photons can be manipulated.\ Therefore the spectral property of this biphoton state can be shaped in either modifying the central frequencies or controlling the phases of the photons \cite{Jen2016b}.\ We further analyze the entropy of entanglement of the multiplexed two-photon source by Schmidt decomposition, which grows as the number of multiplexed ensembles increases.\ 

Here in contrast we propose to generate a multiphoton source out of the cascade emissions from the cascaded cold atomic ensembles.\ We take a two-photon source of quantum-correlated signal and idler photons emitted from a diamond-type atomic configuration, as an initial seed for multiphoton generation in the cascaded atomic ensembles.\ This is motivated by the multiphoton generation from spontaneous parametric down conversion (SPDC) where multiple cascaded nonlinear crystals are pumped sequentially \cite{Anno2006}.\ Similar to the sequential pumping scheme in SPDC, we can take the infrared (telecom) photon of the biphoton source along with a telecom (infrared) driving field to generate another spontaneously emitted cascade emissions satisfying the four-wave mixing condition.\ In this way we enable a $k$-photon source with $m$ infrared and $n$ telecom photons fulfilling $k$ $=$ $m$ $+$ $n$.\ Since the last cascaded atomic ensemble always produce two photons in the infrared and telecom bandwidths respectively, at least one infrared or telecom photon is generated.\ However if the infrared and telecom photons can be converted back and forth with each other \cite{Radnaev2010, Jen2010}, we can have arbitrary $m$ and $n$.\ Furthermore, we investigate its entanglement property in continuous frequency spaces.\ Our studies allow for an alternative setting for multiphoton generation from cold atomic ensembles, and demonstrate spectral shaping to control the entanglement of this multiphoton source \cite{Pan2012}, advancing the development of multimode quantum communication.\ 

In this paper, we first discuss the Hamiltonian and four-wave mixing condition in Sec. II, and propose to generate a multiphoton source in the scheme of cascaded atomic ensembles in Sec. III.\ Then in Sec. IV we study the bipartite entanglement properties of the multiphoton source, and we conclude in Sec. V.

\section{Cascade emissions and four-wave mixing condition}

\begin{figure}[t]
\centering
\includegraphics[width=8.5cm,height=6cm]{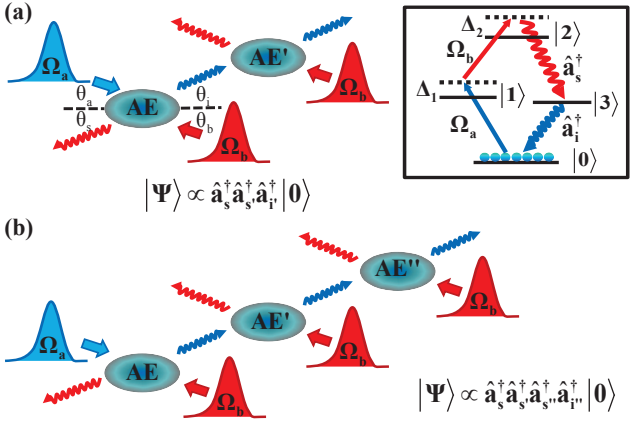}
\caption{(Color online) Proposed multiphoton source from the cascade emissions in the cascaded scheme with a diamond-type atomic configuration.\ In the inset of (a) the signal and idler photon pair $\hat{a}^\dagger_{s,i}$ is created by two pump fields $\Omega_{a,b}$ with single and two-photon detunings $\Delta_{1,2}$, driving the atomic ensembles (AEs).\ For demonstration we plot (a) three and (b) four-photon generations in the cascaded scheme.\ In (a) $\hat{a}^\dagger_{i}$ is sent to AE' along with a pump field $\Omega_b$, and then is converted into a pair of photons $\hat{a}^\dagger_{s'}$ and $\hat{a}^\dagger_{i'}$.\ The excitation and photon emission angles are denoted as $\theta_{a,b,s,i}$ with respect to the long-axis of AE.\ In (b) $\hat{a}^\dagger_{i}$ and $\hat{a}^\dagger_{i'}$ are sent sequentially to AE' and AE'' interacting with pump fields $\Omega_b$, and then effectively are converted into three photons  $\hat{a}^\dagger_{s'}$, $\hat{a}^\dagger_{s''}$, and $\hat{a}^\dagger_{i''}$.\ $|\Psi\rangle$ are the effective multiphoton states in the cascaded scheme.}\label{fig1}
\end{figure}

We consider a Rb atomic ensemble (AE) with a diamond-type configuration shown in the inset of Fig. \ref{fig1}(a).\ The correlated cascade emissions of signal and idler photons ($\hat a_{s(i)}^\dag$ as shorthands for $\hat a_{k_m,\lambda_m}^\dag$ below) are created by pumping the atomic system with two classical fields.\ With dipole approximation of light-matter interactions and rotating-wave approximation (RWA) \cite{QO:Scully}, we express the Hamiltonian in the interaction picture using the same notation of \cite{Jen2016a},
\begin{eqnarray}
V_{\rm I}&=&-\sum_{m=1,2}\Delta_m\sum_{\mu=1}^N|m\rangle_\mu\langle m|-\sum_{m=a,b}\left(\frac{\Omega_m}{2}\hat{P}_m^\dag+{\rm h.c.}\right)\nonumber\\
&&-i\sum_{m=s,i}\bigg\{\sum_{\k_m,\lambda_m}g_m\hat{a}_{\k_m,\lambda_m}\hat{Q}_m^\dag e^{-i\Delta\omega_m t}-{\rm h.c.}\bigg\},\label{H}
\end{eqnarray} 
where we set $\hbar$ $=$ $1$ for simplicity, and denote $\lambda_m$ as the polarizations of photons.\ The collective dipole operators are defined as $\hat{P}_a^\dag$ $\equiv$ $\sum_\mu|1\rangle_\mu\langle 0|e^{i\k_a\cdot\r_\mu}$, $\hat{P}_b^\dag$ $\equiv$ $\sum_\mu|2\rangle_\mu\langle 1|e^{i\k_b\cdot\r_\mu}$, $\hat{Q}_s^\dag$ $\equiv$ $\sum_\mu|2\rangle_\mu\langle 3|e^{i\k_s\cdot\r_\mu}$, and $\hat{Q}_i^\dag$ $\equiv$ $\sum_\mu|3\rangle_\mu\langle 0|e^{i\k_i\cdot\r_\mu}$, with Rabi frequencies $\Omega_{a(b)}$ for two pump fields.\ Central frequencies and wavevectors of these four fields are $\omega_{a(b),s(i)}$ and $\k_{a(b),s(i)}$ respectively.\ Signal and idler photon coupling constants are $g_{s(i)}$ where we have absorbed $(\epsilon_{\k_m,\lambda_m}\cdot\hat{d}_m^*)$ into $g_{s(i)}$ for concise expressions.\ $\epsilon_{\k_m,\lambda_m}$ is the polarization direction of the quantized bosonic fields $\hat{a}_{\k_m,\lambda_m}$, and $\hat{d}_m$ is the unit direction of the dipole operators.\ The detunings are $\Delta_1$ $=$ $\omega_a$ $-$ $\omega_1$ and $\Delta_2$ $=$ $\omega_a$ $+$ $\omega_b$ $-$ $\omega_2$, and for later convenience we define $\Delta\omega_s$ $\equiv$ $\omega_s$ $-$ $\omega_2$ $+$ $\omega_3$ $-$ $\Delta_2$ and $\Delta\omega_i$ $\equiv$ $\omega_i$ $-$ $\omega_3$ with the atomic level energies $\omega_{1,2,3}$.\ The upper level $|2\rangle$ allows for a telecom wavelength within 1.3-1.5 $\mu$m \cite{Chaneliere2006} if 6S$_{1/2}$, 7S$_{1/2}$, or 4D$_{3/2(5/2)}$ levels are considered.

With the Hamiltonian of Eq. (\ref{H}), we can construct self-consistent Schr\"{o}dinger equations assuming there is only one atomic excitation \cite{Jen2012-2, Jen2016a}.\ This is valid when weak and large detuned excitation pulses are considered, satisfying $|\Delta_{1,2}|$ $\gg$ $|\Omega_{a,b}|$.\ After adiabatically eliminating the atomic levels $|1\rangle$ and $|2\rangle$ in the coupled Schr\"{o}dinger equations, we derive the probability amplitude of the biphoton state $|1_{\k_s},1_{\k_i}\rangle$ \cite{Jen2012-2},
\begin{eqnarray}
D_{s,i}(t)&=& g_{i}^{\ast}g_{s}^{\ast}\sum_{\mu=1}^Ne^{i\Delta\k\cdot\r_{\mu}}\int_{-\infty}^{t}\int_{-\infty}^{t^{\prime}}dt^{\prime\prime}dt^{\prime}
e^{i\Delta\omega_{i}t^{\prime}}e^{i\Delta\omega_{s}t^{\prime\prime}}\nonumber\\
&&\times b(t^{\prime\prime})e^{(-\frac{\Gamma_{3}^{ N}}{2}+i\delta\omega_{i})(t^{\prime}-t^{\prime\prime})},\label{Dsi2}
\end{eqnarray}
where $b(t)=\Omega_{a}(t)\Omega_{b}(t)/(4\Delta_{1}\Delta_{2})$, resulting from the required adiabatic conditions for driving pulses.\ Four-wave mixing (FWM) condition $\sum_{\mu=1}^N e^{i\Delta\k\cdot\r_{\mu}}$ represents the phase-matching condition when $\Delta\k$ $=$ $\k_{a}$ $+$ $\k_{b}$ $-$ $\k_{s}$ $-$ $\k_{i}$ $\rightarrow$ $0$ in the limit of large number of atoms $N$, and guarantees to create a highly correlated photon pair.\ Furthermore the idler photon is superradiant \cite{Chaneliere2006, Jen2012, Srivathsan2013} that $\Gamma_{3}^{\rm N}$ $=$ $(N\bar{\mu}+1)\Gamma_{3}$ quantifies a superradiant decay rate in the level of $|3\rangle$ with an intrinsic decay rate $\Gamma_3$.\ The geometrical constant $\bar{\mu}$ \cite{Rehler1971} relates to the shape of the AE, and the associated collective frequency shift \cite{Friedberg1973, Scully2009, Jen2015} is denoted as $\delta\omega_{i}$.\ This reflects the nature of collective radiation \cite{Dicke1954} due to induced dipole-dipole interactions in the dissipation \cite{Lehmberg1970}.\ Note that a complete description of the collective frequency shift requires non-RWA terms in the Hamiltonian \cite{Lehmberg1970, Jen2015}.

This biphoton state $|1_{\k_s},1_{\k_i}\rangle$ has two main features.\ One is the strong directionality of the emitted photons, which is determined by FWM.\ In thermodynamic limit of AE, we have $\Delta\k$ $=$ $0$.\ If counter-propagating excitations \cite{Chaneliere2006} are used as in Fig. \ref{fig1}(a) with excitation and photon emission angles denoted as $\theta_{a,b,s,i}$, we can derive the condition for the emitted angles as 
\bea
&&\frac{\lambda_b}{\lambda_a}(\cos\theta_a-\cos\theta_i)=\cos\theta_b-\cos\theta_s,\nonumber\\
&&\frac{\lambda_b}{\lambda_a}(\sin\theta_a+\sin\theta_i)=\sin\theta_b+\sin\theta_s,\label{FWM}
\eea
where $\lambda_{a,b}$ are the wavelengths of the excitation fields, and we have assumed $|\k_i|$ $=$ $|\k_a|$ and $|\k_s|$ $=$ $|\k_b|$.\ Since $\theta_{a,b}$ are given when the excitations are applied, $\theta_{s,i}$ can be decided from Eq. (\ref{FWM}).\ To have some estimate for the angles, we have ($\theta_i,\theta_s$) $=$ ($4.9^{\circ},9.9^{\circ}$) when we set ($\theta_a,\theta_b$) $=$ $(5^{\circ},10^{\circ})$ and $\lambda_b/\lambda_a$ $=$ $2$.\ For the same ratio of the wavelengths, we have ($\theta_i,\theta_s$) $=$ ($7.9^{\circ},13.9^{\circ}$) when we set ($\theta_a,\theta_b$) $=$ $(4^{\circ},10^{\circ})$.\ If in radians that $\theta_{a,b,s,i}$ $\ll$ $1$ and setting $\theta_b$ $=$ $2\theta_a$, we derive $\theta_s$ ($\approx$ $\theta_b$) $\approx$ $2\theta_i$, which indicates that the signal and idler photons are emitted with corresponding excitation angles of $\theta_b$ and $\theta_a$ respectively, and they follow the directions almost tangent to the long axis of AE.\ This direction is preferential in experiments, which allows for strong light-matter couplings.\ 

The other feature is that the biphoton state is probabilistic.\ According to Eq. (\ref{Dsi2}), we can express the complete and normalized state as
\bea
|\Psi\rangle=\frac{1}{\sqrt{1+|D_{s,i}|^2}}|0^{\otimes N}\rangle\left[|{\rm vac}\rangle + D_{s,i}|1_{\k_s},1_{\k_i}\rangle\right],
\eea
which involves $N$ atomic ground with vacuum and biphoton states respectively.\ Since $|D_{s,i}|$ $\ll$ $1$, most of the time AE does not generate any photons.\ Therefore experimentally it requires repeated excitations until the biphoton state is created, which can be confirmed via photon detections.\ The degree of correlation of the photons (second-order correlation function for example) can be measured as well by a conditional detection \cite{Chaneliere2006, Srivathsan2013} of the idler photon after the signal one is detected.\ Below and throughout the paper, we focus on the effective biphoton and multiphoton states, therefore the normalization for these effective states is neglected.\ Due to their probabilistic feature, these effective states can be confirmed only via conditional detections or post selections.

Specifically we use normalized Gaussian pulses where $\Omega_{a(b)}(t)$ $=$ $[\sqrt{\pi}\tau_{a(b)}]^{-1}\tilde{\Omega}_{a(b)}e^{-t^{2}/\tau_{a(b)}^{2}}$ with the pulse areas $\tilde{\Omega}_{a,b}$ and pulse widths $\tau_{a(b)}$.\ Considering the long time limit that $D_{s,i}(t\rightarrow\infty)$, we derive the probability amplitude $D_{si}$ after inserting the Gaussian forms into Eq. (\ref{Dsi2}) and redefining $\Delta\omega_{s(i)}$ as $\Delta\omega_{s}-\delta\omega_i$ and $\Delta\omega_{i}+\delta\omega_i$ respectively, which reads
\begin{eqnarray}
D_{si}(\Delta\omega_s,\Delta\omega_i)=\frac{\tilde{\Omega}_a\tilde{\Omega}_b g_s^*g_i^*\sum_{\mu=1}^N e^{i\Delta\k\cdot\r_\mu}}{4\Delta_1\Delta_2\sqrt{2\pi}\sqrt{\tau_a^2+\tau_b^2}}f(\omega_s,\omega_i).\label{Dsi}
\end{eqnarray}
The spectral function of the biphoton state is
\begin{eqnarray}
f(\omega_s,\omega_i)=\frac{e^{-(\Delta\omega_s+\Delta\omega_i)^2\tau_{eff}^2/8}}{\frac{\Gamma_3^{ N}}{2}-i\Delta\omega_i},\label{f}
\end{eqnarray}
where $\tau_{eff}$ $\equiv$ $\sqrt{2}\tau_a\tau_b/\sqrt{\tau_a^2+\tau_b^2}$.\ This spectrally correlated biphoton state involves a Gaussian weighting modulated by a Lorentzian.\ The spectral function is most significant when $\Delta\omega_s$ $\approx$ $-\Delta\omega_i$ within the spectral range of $1/\tau_{eff}$.\ More entangled biphoton state can be made with an increasing optical density of the AE or longer pulses \cite{Jen2012-2}, making $f(\omega_s,\omega_i)$ less factorizable aligning on the axis of $\Delta\omega_s$ $=$ $-\Delta\omega_i$ \cite{Jen2016a}.\ In the next section we propose to use this two-photon source as a seed to create multiphoton states in the scheme of cascaded AEs.

\section{Multiphoton states from the cascade emissions in the cascaded scheme}

Before we investigate the multiphoton states generated from the cascaded atomic ensembles, we note that throughout the paper, we focus only on the effective pure states generation.\ In general for an open quantum system, the system becomes mixed states inevitably \cite{Gardiner2015} due to the interactions with the reservoir or the imperfections in experiments.\ The mixed states can be interpreted as statistical mixtures of pure states, which can be expressed as a density operator, $\hat \rho$ $=$ $\sum_k p_k|\psi_k\rangle\langle\psi_k|$ with the constraints of $\sum_kp_k$ $=$ $1$ and $0$ $<$ $p_k$ $\leq$ $1$.\ $\hat \rho$ becomes a pure state when one of $p_k$'s is unity, therefore this density operator shows a general representation of any quantum system.\ For example of the spontaneous emission process in a two-level atom ($|0\rangle$ and $|1\rangle$), it acts like an amplitude damping \cite{Nielsen2000} to the atomic states.\ In a description of generalized amplitude damping, the atom evolves to the stationary mixed state $\rho_{\infty}$ $=$ $p|0\rangle\langle 0|$ $+$ $(1-p)|1\rangle\langle 1|$ with some probability $p$.\ Similarly, $\rho_\infty$ can describe a loss of photon due to attenuation or decoherence from environment if $|0\rangle$ and $|1\rangle$ are denoted as vacuum and one-photon states respectively.\ 

In general, a spontaneous emission is a random process \cite{QO:Scully, Nielsen2000} in time and space, where the emitted direction has a uniformly $d\hat \Omega/(4\pi)$ distribution in a solid angle of $d\hat\Omega$ $=$ $\sin\theta d\theta d\phi$ in spherical coordinates.\ In contrast to this random process in space, the biphoton state from the cascade emissions in a diamond configuration under FWM condition is highly directional and correlated, which thus makes the effective state $|1_{\k_s},1_{\k_i}\rangle$ valid if photon loss or other noisy channels that deteriorate its fidelity can be neglected.\ Although this limits our investigation to the pure states, we note that an entanglement distillation or purification procedure \cite{Bennett1996L, Nielsen2000, Braunstein2005} can be applied to the mixed entangled states by local operations and classical communication.\ Therefore the pure state picture we focus here can still give insights to quantum information processing with the effective multiphoton state presented in this work.\ Other discussions of inseparability criterion, bound entanglement, or multipartite entanglement for mixed states can be referred to Refs. \cite{Braunstein2005, Pan2012}.\ For measuring entanglement, it does not imply Bell nonlocality except for pure states, and quantifying entanglement for bipartite mixed state involves various measures not agreeable to the partial von Neumann entropy \cite{Braunstein2005}.\ That being the case, our investigations using Schmidt decomposition to quantify the entanglement of bipartite pure states in Sec. IV provide an upper bound to the entanglement measure.

To generate a multiphoton state from the cascade emissions, we propose to couple one of the two photons with other AE along with a corresponding pump field, such that $(k+1)$-photon source can be created from a $k$-photon state.\ As demonstrated in Fig. \ref{fig1}, three and four-photon states are created in the cascaded scheme.\ AE' and AE'' are two other atomic ensembles used to generate correlated photons $\hat a_{s',i'}^{\dag}$ and $\hat a_{s'',i''}^{\dag}$ respectively under the FWM condition.\ Since the initial seed of two-photon source is highly entangled in frequency space, the generated multiphoton source is expected to be also entangled in continuous frequency spaces.\ 

The generated multiphoton spectral functions can be derived by products of the two-photon spectral ones of Eq. (\ref{f}) and invoke the condition for central frequencies of the annihilated photon, that is $\omega_i$ $=$ $\omega_s'$ $+$ $\omega_i'$ $-$ $\omega_b$ for example in the case of AE' in Fig. \ref{fig1}(a).\ This condition also satisfies the energy conservation as if four fields are plane waves.\ Below we demonstrate three and four-photon state spectral functions, and their entanglement properties will be discussed in the next section.

\subsection{Three-photon state}

As in Fig. \ref{fig1}(a) which we denote a route $B1$ for the three-photon generation, the effective three-photon state involving two signal and one idler photons can be expressed as
\bea
|\Psi\rangle_{3,B1}=f_{3,B1}\hat{a}^\dag_s\hat a_{s'}^\dag\hat a_{i'}^\dag|0\rangle.
\eea
We derive the dimensionless and effective spectral function by multiplying a typical biphoton state spectral function of Eq. (\ref{f}) with the one generated by a plane-wave idler photon $\hat a_i^\dag$ and a driving field $\Omega_b$ in AE',
\bea
f_{3,B1}/\Gamma_3^2=\frac{e^{-(\Delta\omega_s+\Delta\omega_i)^2\tau_{eff}^2/8}}{\frac{\Gamma_3^{ N}}{2}-i\Delta\omega_i}
\frac{e^{-(\Delta\omega_{s'}+\Delta\omega_{i'})^2\tau_b^2/4}}{\frac{\Gamma_3^{ N}}{2}-i\Delta\omega_{i'}},
\eea
where $\tau_b$ appears when we let $\tau_a$ $\rightarrow$ $\infty$ in $\tau_{eff}$.\ The above becomes, after setting $\omega_i$ $=$ $\omega_s'$ $+$ $\omega_i'$ $-$ $\omega_b$ and using $\Delta_2$ $=$ $\omega_a$ $+$ $\omega_b$ $-$ $\omega_2$,
\bea
f_{3,B1}/\Gamma_3^2&=&\frac{e^{-(\Delta\omega_s+\Delta\omega_{s'}+\Delta\omega_{i'}+\Delta_{a3}+\delta\omega_i)^2\tau_{eff}^2/8}}{\frac{\Gamma_3^{ N}}{2}-i(\Delta\omega_{s'}+\Delta\omega_{i'}+\Delta_{a3}+\delta\omega_i)}\nonumber\\
&\times&\frac{e^{-(\Delta\omega_{s'}+\Delta\omega_{i'})^2\tau_b^2/4}}{\frac{\Gamma_3^{ N}}{2}-i\Delta\omega_{i'}}.\label{f3}
\eea
The extra frequency shift $\Delta_{a3}$ $\equiv$ $\omega_a$ $-$ $\omega_3$ in the above can be removed along with $\delta\omega_i$ by applying an external Zeeman field, such that we can simplify Eq. (\ref{f3}) as
\bea
\frac{f_{3,B1}}{\Gamma_3^2}=\frac{e^{-(\Delta\omega_s+\Delta\omega_{s'}+\Delta\omega_{i'})^2\tau_{eff}^2/8}}{\frac{\Gamma_3^{ N}}{2}-i(\Delta\omega_{s'}+\Delta\omega_{i'})}\frac{e^{-(\Delta\omega_{s'}+\Delta\omega_{i'})^2\tau_b^2/4}}{\frac{\Gamma_3^{ N}}{2}-i\Delta\omega_{i'}}.\label{f3-1}\nonumber\\
\eea

Similarly, if we annihilate signal photon of the initial two-photon seed, we have alternatively the effective three-photon state as ($B2$ to denote the second route for three-photon state generation, involving two idler and one signal photons)
\bea
|\Psi\rangle_{3,B2}=f_{3,B2}\hat{a}^\dag_i\hat a_{s'}^\dag\hat a_{i'}^\dag|0\rangle,
\eea
where its dimensionless spectral function is [after removing the extra frequency shift as in deriving Eq. (\ref{f3-1})]
\bea
\frac{f_{3,B2}}{\Gamma_3^2}&=&\frac{e^{-(\Delta\omega_i+\Delta\omega_{s'}+\Delta\omega_{i'})^2\tau_{eff}^2/8}e^{-(\Delta\omega_{s'}+\Delta\omega_{i'})^2\tau_a^2/4}}{(\frac{\Gamma_3^{ N}}{2}-i\Delta\omega_{i})(\frac{\Gamma_3^{ N}}{2}-i\Delta\omega_{i'})}.\label{f3-2}\nonumber\\
\eea
Note that the overall constants in spectral functions do not make an effect on spectral distributions, thus we regularize them in dimensionless forms.\ The overall constants however determine the generation rates which are small in general since weak and large detuned excitations are used.\

Eqs. (\ref{f3-1}) and (\ref{f3-2}) are two of the main results in this subsection.\ Obviously these three-photon states are entangled in frequency spaces, meaning there is no possible ways to factorize these spectral functions.\ Meanwhile $|\Psi\rangle_{3,B1}$ differs from $|\Psi\rangle_{3,B2}$ specifically in a modulated Lorentzian function on signal distribution $\Delta\omega_{s'}$.\ This results from the annihilation of the idler photon in the route $B1$, which replaces the Lorentzian with the correlated signal and idler photon pair.\ Also different timescales $\tau_{b(a)}$ in routes $B1(2)$ respectively for a joint Gaussian profile of the photon pair $\hat a_{s'}^\dag\hat a_{i'}^\dag$ suggest an independent control over their spectral functions by varying pulse durations.

Below we derive the spectral function for the four-photon state of Fig. \ref{fig1}(b) using the three-photon states in the cascaded scheme of Fig. \ref{fig1}(a), and also the other possible spectral functions in alternative routes.

\subsection{Four-photon state}

Here using the same fashion to generate three-photon states, the four-photon ones can be also created in the cascaded scheme.\ As in Fig. \ref{fig1}(b), the idler photon $\hat a_{i'}^\dag$ emitted from AE' is annihilated with an extra coupling field $\Omega_b$ in AE'' to generate a newly correlated pair of photons $\hat a_{s''}^\dag \hat a_{i''}^\dag$.\ The effective four-photon state with three signal and one idler photons becomes
\bea
|\Psi\rangle_{4,C1}=f_{4,C1}\hat{a}^\dag_s\hat a_{s'}^\dag\hat a_{s''}^\dag\hat a_{i''}^\dag|0\rangle,
\eea
where again the dimensionless spectral function can be derived as 
\begin{widetext}
\bea
\frac{f_{4,C1}}{\Gamma_3^3}=\frac{e^{-(\Delta\omega_s+\Delta\omega_{s'}+\Delta\omega_{s''}+\Delta\omega_{i''})^2\tau_{eff}^2/8}e^{-(\Delta\omega_{s'}+\Delta\omega_{s''}+\Delta\omega_{i''})^2\tau_b^2/4}e^{-(\Delta\omega_{s''}+\Delta\omega_{i''})^2\tau_b^2/4}}
{\big[\frac{\Gamma_3^{ N}}{2}-i(\Delta\omega_{s'}+\Delta\omega_{s''}+\Delta\omega_{i''})\big]\big[\frac{\Gamma_3^{ N}}{2}-i(\Delta\omega_{s''}+\Delta\omega_{i''})\big](\frac{\Gamma_3^{ N}}{2}-i\Delta\omega_{i''})}.\label{f4-1}
\eea
\end{widetext}

Other four possible routes to generate four-photon states are demonstrated in Appendix A.\ In principle there should be six different spectral functions where three of them are from routes $B1$ and $B2$ respectively.\ Since there is one spectral function which is symmetric to each other in respective routes, making a total of five possible spectral functions in our cascaded scheme.\ This symmetric four-photon state is generated by annihilating $\hat a_{s(i)}^\dag$ from the routes $B1(2)$ respectively along with the coupling fields $\Omega_{a(b)}$.\ The symmetry is satisfied when $\hat a_{s''}$ $\leftrightarrow$ $\hat a_{s'}$ and $\hat a_{i''}$ $\leftrightarrow$ $\hat a_{i'}$, and this effective four-photon state with two signal and two idler photons is
\bea
|\Psi\rangle_{4,C3}=f_{4,C3}\hat{a}^\dag_{s'}\hat a_{i'}^\dag\hat a_{s''}^\dag\hat a_{i''}^\dag|0\rangle,
\eea
with the spectral function
\begin{widetext}
\bea
\frac{f_{4,C3}}{\Gamma_3^3}=\frac{e^{-(\Delta\omega_{s'}+\Delta\omega_{i'}+\Delta\omega_{s''}+\Delta\omega_{i''})^2\tau_{eff}^2/8}e^{-(\Delta\omega_{s'}+\Delta\omega_{i'})^2\tau_a^2/4}e^{-(\Delta\omega_{s''}+\Delta\omega_{i''})^2\tau_b^2/4}}
{\big[\frac{\Gamma_3^{ N}}{2}-i(\Delta\omega_{s''}+\Delta\omega_{i''})\big](\frac{\Gamma_3^{ N}}{2}-i\Delta\omega_{i'})(\frac{\Gamma_3^{ N}}{2}-i\Delta\omega_{i''})}.\label{C3}
\eea
\end{widetext}

These spectral functions have a common weighting of Gaussian envelope involving four photon frequencies, indicating to possess a genuine $k$-party entanglement \cite{Braunstein2005} for $k$-photon source in our proposed cascaded scheme.\ This genuine multiphoton entanglement means to exclude any possible bipartite splittings or groupings \cite{Braunstein2005}.\ For example of Eq. (\ref{C3}) from the route $C3$, two groups of photons $\hat a_{s',i'}^\dag$ and $\hat a_{s'',i''}^\dag$ respectively would be able to be factorized if this common weighting of $e^{-(\Delta\omega_{s'}+\Delta\omega_{i'}+\Delta\omega_{s''}+\Delta\omega_{i''})^2\tau_{eff}^2/8}$ is absent.\ Multipartite entanglement is still an ongoing research even for pure states we consider here, therefore to give an intuitive study of entanglement property of our proposed multiphoton sources, in the next section we introduce Schmidt decomposition to investigate their bipartite entanglements in continuous frequency spaces.

\section{Entropy of entanglement}

To gain insights of the spectral entanglement in the multiphoton sources from the cascaded scheme of AEs, we study the bipartite entropy of entanglement in these sources.\ The multiphoton states can be projected or collapsed to the effective biphoton state by conditional measurements, such that we can quantify and analyze their spectral entanglement by Schmidt decomposition \cite{Law2000}.\ The bipartite entropy of entanglement is $S$ $=$ $-$ $\sum_{n=1}^\infty$ $\lambda_n \rm{log}_2\lambda_n$ \cite{Bennett1996}, intended for pure states.\ Schmidt eigenvalues $\lambda_n$ determine the probabilities of $n$th mode functions, with which the state can be expressed as $\sum_{n}\sqrt{\lambda_n}\hat{b}_n^\dag\hat{c}_n^\dag|0\rangle$, for two effective photon operators $\hat{b}_n$ and $\hat{c}_n$ associated with mode functions $\psi_n$ and $\phi_n$ respectively.\ 

Schmidt decomposition is done numerically as solving the eigenvalue problems of the one-photon spectral kernels \cite{Law2000, Parker2000}, 
\bea
\int K_{1}(\omega,\omega^{\prime})\psi_{n}(\omega^{\prime})d\omega^{\prime}  &=&\lambda_{n}\psi_{n}(\omega),\\
\int K_{2}(\omega,\omega^{\prime})\phi_{n}(\omega^{\prime})d\omega^{\prime}  &=&\lambda_{n}\phi_{n}(\omega),
\eea
where 
\bea
K_{1}(\omega,\omega^{\prime}) &\equiv&\int f(\omega,\omega_{1})f^{\ast}(\omega^{\prime},\omega_{1})d\omega_{1}, \\
K_{2}(\omega,\omega^{\prime}) &\equiv&\int f(\omega_{2},\omega)f^{\ast}(\omega_{2},\omega^{\prime})d\omega_{2}. 
\eea
The spectral function $f$ comes from the effective biphoton state (a pair of $\hat a_s^\dag$ and $\hat a_i^\dag$ for example)
\bea
|\Psi\rangle &=& \int f(\omega_s,\omega_i)\hat{a}_{s}^\dag(\omega_s)\hat{a}_{i}^\dag(\omega_i)|0\rangle d\omega_s d\omega_i,\\
&=&\sum_{n}\sqrt{\lambda_n}\hat{b}_n^\dag\hat{c}_n^\dag|0\rangle,
\eea
with two mode functions in the Schmidt bases, determining the effective photon operators,
\bea
\hat{b}_n^\dag&\equiv&\int\psi_n(\omega_s)\hat{a}_{s}^\dag(\omega_s)d\omega_s,\\
\hat{c}_n^\dag&\equiv&\int\phi_n(\omega_i)\hat{a}_{i}^\dag(\omega_i)d\omega_i.
\eea
The above decomposition has been used to quantify the spectral entanglement of two-photon source from parametric down conversion \cite{Law2000, Parker2000} and cascade emissions \cite{Jen2012-2}, and in spectral shaping of biphoton state in a multiplexed scheme \cite{Jen2016a, Jen2016b}.

\subsection{\texorpdfstring{$\boldsymbol{S}$}{S} in three-photon states}

\begin{figure}[t]
\centering
\includegraphics[width=8.5cm,height=4.0cm]{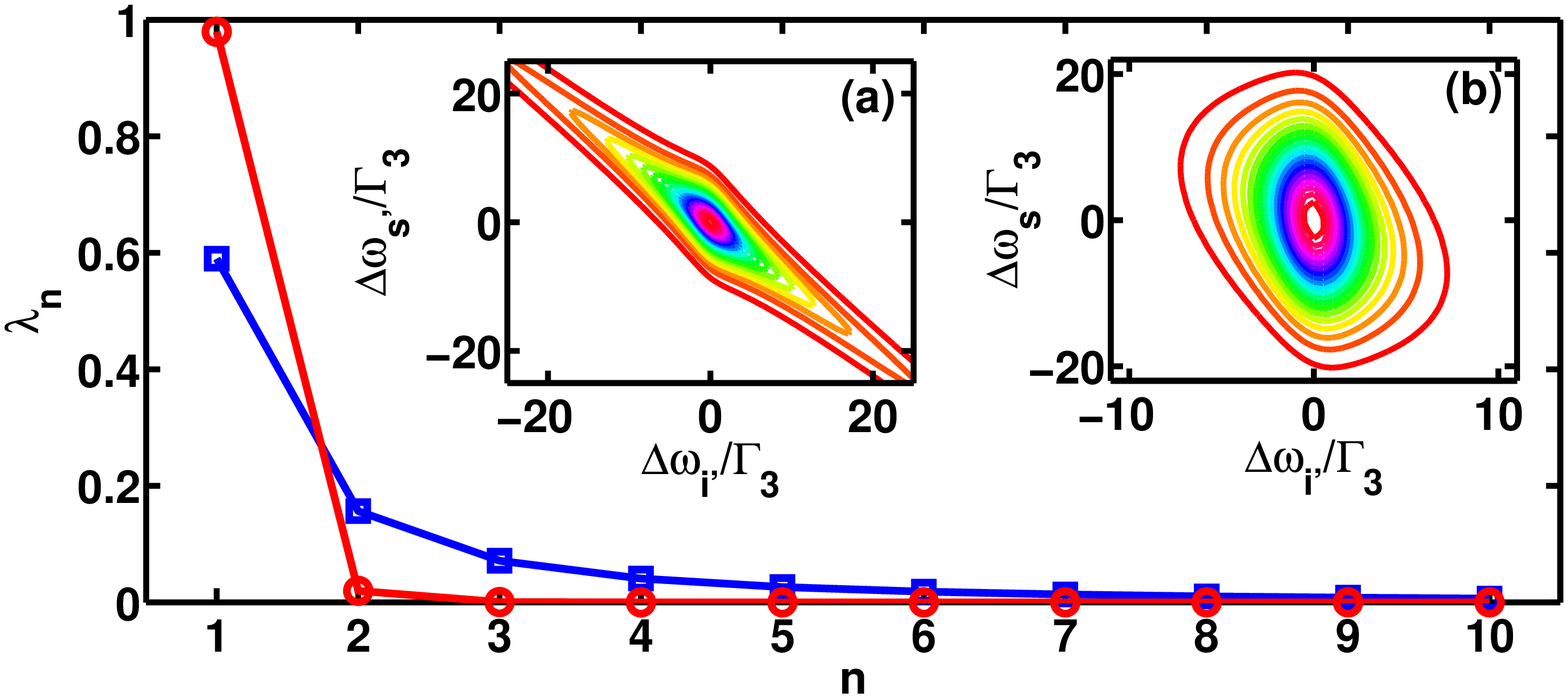}
\caption{(Color online) First ten Schmidt eigenvalues $\lambda_n$ of the projected spectral distributions $|f_{3, B1}|$.\ The spectral ranges for both signal and idler photons are post-selected to $\pm$ $200\Gamma_3$ throughout all figures where we also set $\Gamma^{\rm N}_3$ $=$ $5\Gamma_3$ and $\tau_a$ $=$ $\tau_b$ $=$ $0.25\Gamma_3^{-1}$ without loss of generality.\ The insets (a) and (b) are projected spectral distributions with $\Delta\omega_{s(s')}$ $=$ $0$ respectively with corresponding Schmidt eigenvalues ($\square$) and ($\circ$).}\label{fig2}
\end{figure}

For bipartite entanglement property of a three-photon state with the spectral function $f_{3,B1}$, three possible projected biphoton states can be derived by either annihilating photons $\hat{a}_s$, $\hat a_{s'}$, or $\hat a_{i'}$.\ In the insets of (a) and (b) of Fig. \ref{fig2}, we show the spectral distributions projecting out $\hat a_s$ and $\hat a_{s'}$ respectively by setting $\Delta\omega_{s(s')}$ $=$ $0$.\ The entropy of entanglement $S$ is $2.37$ and $0.15$ respectively, indicating a more entangled source in the former case.\ The huge difference in these two projections results from a removal of an entangling function $e^{-(\Delta\omega_{s'}+\Delta\omega_{i'})^2\tau_b^2/4}$ of Eq. (\ref{f3-1}) when projecting out $\hat a_{s'}$.\ This Gaussian function supposes to entangle photons $\hat a_{s'}$ and $\hat a_{i'}$ in infinite signal and idler frequency spaces.\ 
\begin{figure}[b]
\centering
\includegraphics[width=8.5cm,height=4.0cm]{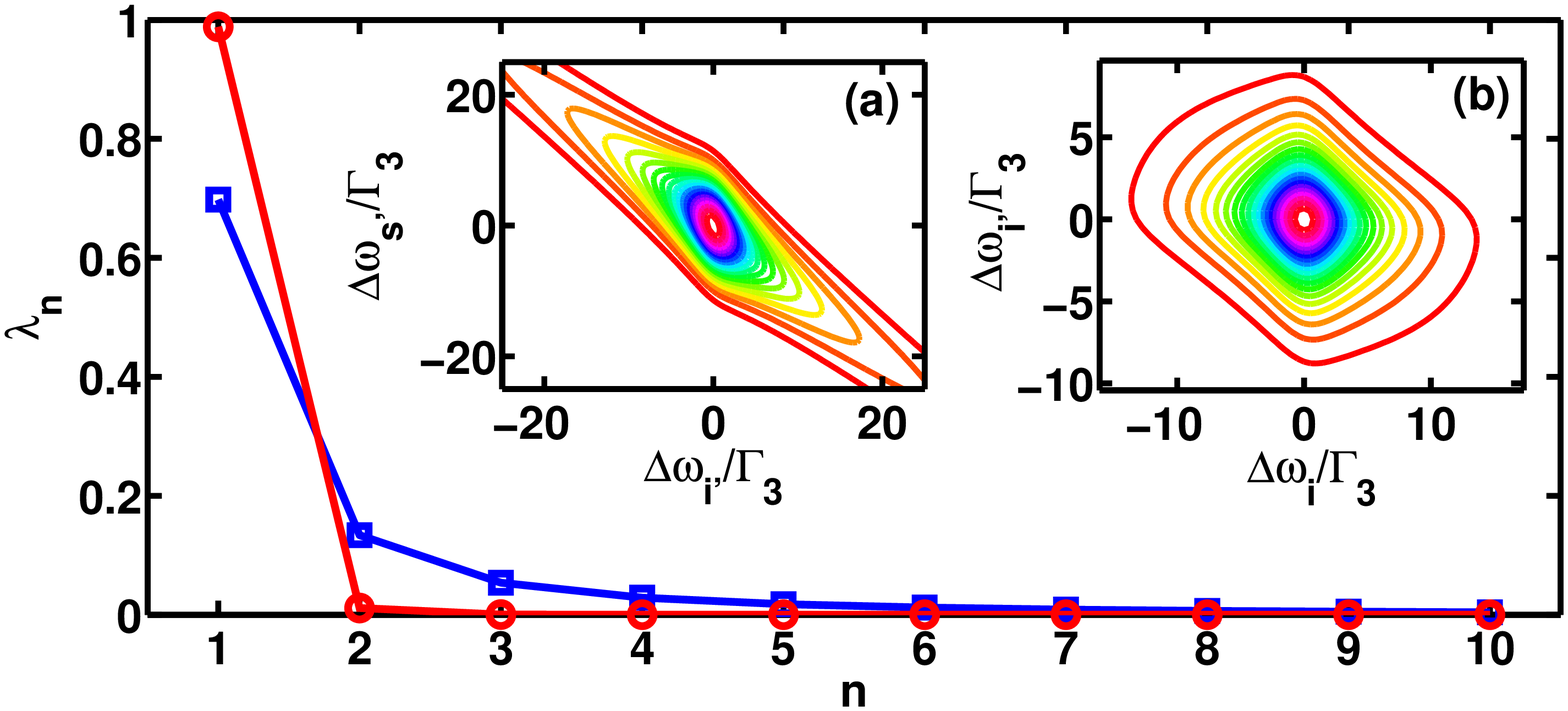}
\caption{(Color online) First ten Schmidt eigenvalues $\lambda_n$ of the projected spectral distributions $|f_{3, B2}|$.\ The parameters are the same as in Fig. \ref{fig2} while the insets (a) and (b) are projected spectral distributions with $\Delta\omega_{i(s')}$ $=$ $0$ and corresponding Schmidt eigenvalues ($\square$) and ($\circ$) respectively.}\label{fig3}
\end{figure}
A modulating Lorentzian function of idler photon however confines its spectral distribution into a finite bandwidth of $\sim$ $\Gamma_3^N$.\ Projecting out a signal photon $\hat a_{s'}$ thus collapses the entangling function into $e^{-(\Delta\omega_{i'})^2\tau_b^2/4}$, rendering another finite bandwidth of the idler photon with FWHM (full-width at half-maximum) $\sim$ $4\sqrt{\ln 2}/\tau_b$.\ In essence these factorizable idler functions, Gaussian and Lorentzian, rotate the spectral distribution toward the axis $\Delta\omega_{i'}$ $=$ $0$, as we show in Fig. \ref{fig2}(b).\ The third possible projected biphoton state is to project out $\hat a_{i'}$, which has a similar spectral property to the one annihilating $\hat a_{s'}$.\

Likewise in Fig. \ref{fig3}, we show the results for the spectral function $f_{3,B2}$.\ The insets of (a) and (b) demonstrate the spectral functions of the projected biphoton states with annihilated photons $\hat{a}_i$ and $\hat a_{s'}$ respectively.\ The entropy of entanglement $S$ is $1.79$ and $0.09$ respectively, again indicating a more entangled source in the former case.\ Fig. \ref{fig3}(b) is less entangled due to a removal of the signal photon dependence $\Delta\omega_{s'}$ in the entangling function $e^{-(\Delta\omega_{s'}+\Delta\omega_{i'})^2\tau_a^2/4}$, similar to the cases in Fig. \ref{fig2}.\ The reason why Fig. \ref{fig3}(a) is less entangled than Fig. \ref{fig2}(a) is that an extra entangling Lorentzian function $[\Gamma_3^{ N}/2-i(\Delta\omega_{s'}+\Delta\omega_{i'})]^{-1}$ present in Eq. (\ref{f3-1}).\ Fig. \ref{fig3}(b) shows a relatively smaller entangled biphoton source, resulting from a spectral function with more aligned distributions on $\Delta\omega_{i(i')}$ $=$ $0$ compared to Fig. \ref{fig2}(b), in a somewhat distorted fashion of Fig. \ref{fig2}(b).\ This is due to two factorizable Lorentzian functions of idler photons $\hat a_{i}$ and $\hat a_{i'}$ in frequencies $\Delta\omega_{i(i')}$ of Eq. (\ref{f3-2}) when projecting out the signal photon $\hat a_{s'}$.\

\begin{figure}[b]
\centering
\includegraphics[width=8.5cm,height=4.0cm]{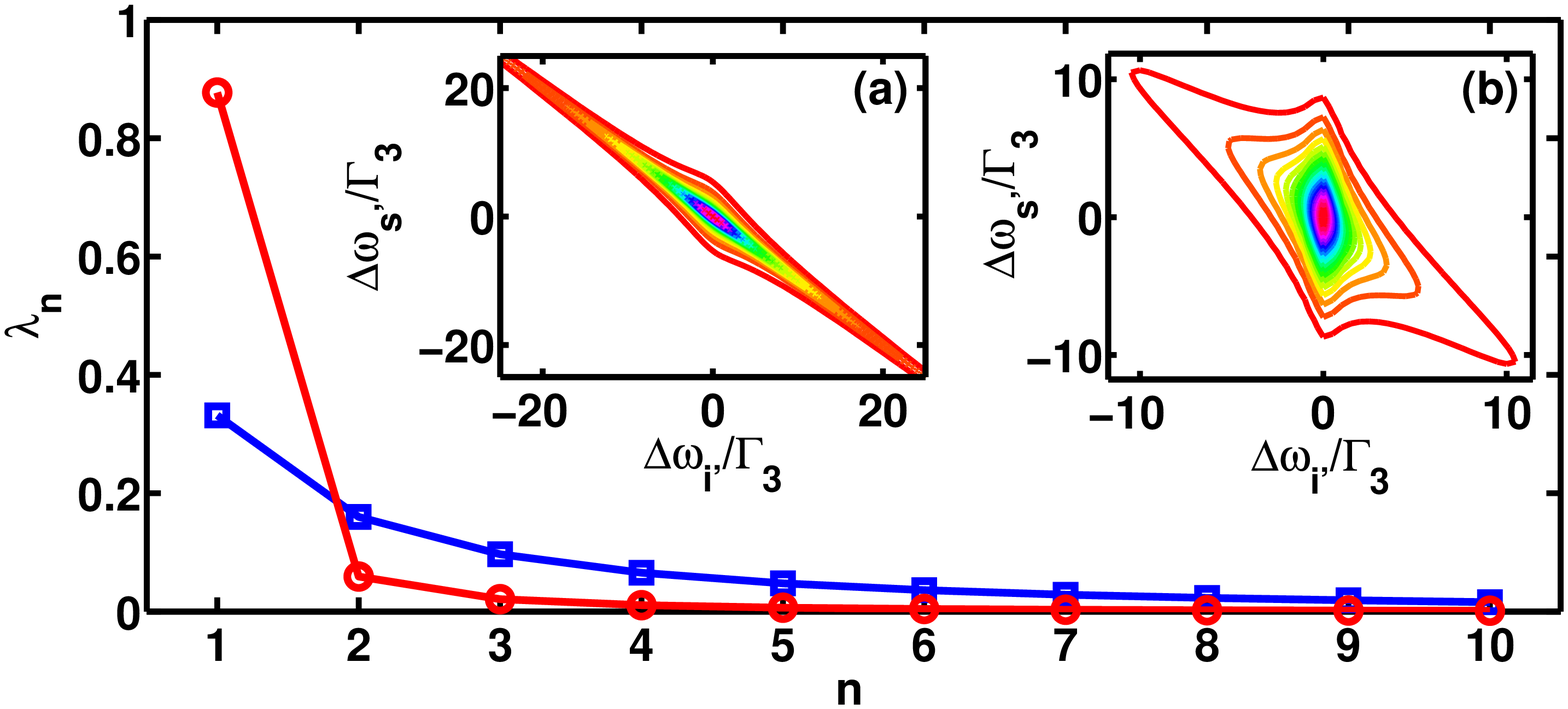}
\caption{(Color online) First ten Schmidt eigenvalues $\lambda_n$ of the projected spectral distributions $|f_{3, B1}|$ with different decay constants $\Gamma_3^N$ and $\Gamma_3^{N'}$.\ The parameters are the same as in Fig. \ref{fig2} while the insets (a) and (b) are projected spectral distributions as in Fig \ref{fig2}(a) with $\Gamma_3^N/\Gamma_3$ $=$ $1(5)$ and $\Gamma_3^{N'}/\Gamma_3$ $=$ $5(1)$ respectively.\ Corresponding Schmidt eigenvalues are ($\square$) and ($\circ$).}\label{fig4}
\end{figure}

Since the decay constants of two atomic ensembles, AE and AE', are not necessarily the same, we further investigate the spectral property with different $\Gamma_3^N$ and $\Gamma_3^{N'}$.\ As an example from Eq. (\ref{f3-1}) and projecting out $\hat a_{s}$, we have
\bea
\frac{f_{3\rightarrow 2,B1}}{\Gamma_3^2}=\frac{e^{-(\Delta\omega_{s'}+\Delta\omega_{i'})^2\tau_{eff}^2/8}}{\frac{\Gamma_3^{ N}}{2}-i(\Delta\omega_{s'}+\Delta\omega_{i'})}\frac{e^{-(\Delta\omega_{s'}+\Delta\omega_{i'})^2\tau_b^2/4}}{\frac{\Gamma_3^{N'}}{2}-i\Delta\omega_{i'}}.\nonumber\\
\eea
In Fig. \ref{fig4} we show the results for different decay constants compared to Fig \ref{fig2}(a).\ The entropy of entanglement $S$ for the insets (a) and (b) are $3.9$ and $0.89$ respectively.\ The small $\Gamma_3^N$ in (a) provides a sharp distribution along the highly entangled axis $\Delta\omega_{s'}$ $=$ $-\Delta\omega_{i'}$, therefore making this projected spectral function more entangled.\ In contrast the small $\Gamma_3^{N'}$ in (b) limits the factorizable Lorentzian idler distribution $\hat a_{i'}$, allowing for a squeezed distribution in $\Delta\omega_{i'}$ and a less entangled biphoton source.\ 

In this subsection we have demonstrated a rich spectral property of the projected biphoton state from the three-photon source in the cascaded scheme.\ The entanglement property can be very different depending on how the counterpart of the source is collapsed.\ In the perspective of generating a more entangled photon source, the spectral function from route $B1$ serves better than $B2$, meanwhile a less(more) superradiant decay constant $\Gamma_3^N(\Gamma_3^{N'})$ of AE(AE') in the route $B1$ is favorable for this purpose.

\subsection{\texorpdfstring{$\boldsymbol{S}$}{S} in four-photon states}

\begin{figure}[t]
\centering
\includegraphics[width=8.5cm,height=3.8cm]{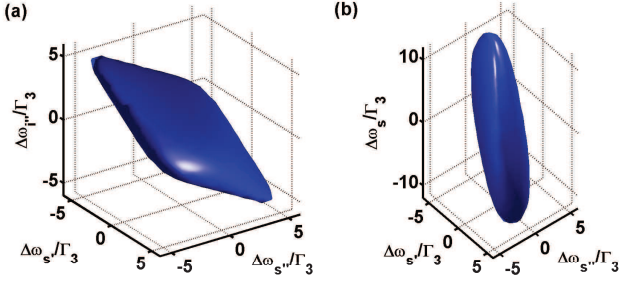}
\caption{(Color online) Three-dimensional isosurface plots for projected spectral distributions $|f_{4,C1}|$ with annihilated photon (a) $\hat a_s$ or (b) $\hat a_{i''}$.\ The isosurface plots are chosen for respectively projected and normalized $|f_{4,C1}|$ $=$ $0.3$ where they show tilted and axial spectral distributions in (a) and (b) respectively.}\label{fig5}
\end{figure}

For four-photon states, there are in general six possible projections to biphoton ones.\ Though of plenty of possible projected biphoton states for a total of five spectral functions demonstrated in Sec. III.B and Appendix A, there are only a few of qualitatively different spectral functions.\ As an example, we choose the spectral function $f_{4,C1}$ of Eq. (\ref{f4-1}).\ Based on the observation from projected three-photon states in the previous subsection, we expect that a projection of $\hat a_{i''}$ or $\hat a_{s''}$ in $f_{4,C1}$ provides less entangled multiphoton states.\ For comparisons, in Fig. \ref{fig5} we show two projections of $f_{4,C1}$ in three-dimensional isosurface plots which provide a qualitative distinction in the projected spectral functions.\ In Fig. \ref{fig5} (a) and (b) we set $\Delta\omega_{s(i'')}$ $=$ $0$ respectively, which show a tilted and an axial distribution.\ The tilted distribution potentially allows for a more entangled multiphoton state while we note that further projecting out photons $\hat a_{i''}$ and $\hat a_s$ respectively in (a) and (b) gives the same spectral function of the projected biphoton states.

\begin{figure}[t]
\centering
\includegraphics[width=8.5cm,height=4.0cm]{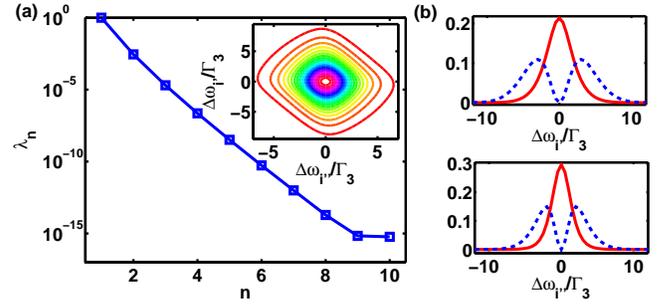}
\caption{(Color online) First ten Schmidt eigenvalues $\lambda_n$ and two mode functions of the projected spectral distributions $|f_{4, B3}|$ with annihilated photons $\hat a_{s'}$ and $\hat a_{s''}$.\ The parameters are the same as in Fig. \ref{fig2}.\ (a) $\lambda_n$ shows an abrupt decrease in logarithmic scale, indicating an extremely low entropy of entanglement.\ (b) Two idler mode probability densities (solid and dashed for the first and the second modes) in $\Delta\omega_{i'(i'')}$ respectively.}\label{fig6}
\end{figure}

Less entangled biphoton states from $f_{4,C1}$ can be derived by projecting out a pair of photons ($\hat a_{s''(i'')}$, $\hat a_{s'}$) or ($\hat a_{s''}$, $\hat a_{i''}$).\ These projections basically collapse $f_{4,C1}$ into a single Gaussian function of $e^{-[\Delta\omega_{s''(i'')}]^2\tau_b^2/4}$ or $e^{-(\Delta\omega_{s'})^2\tau_b^2/4}$, which again confines its spectral distribution without entangling with other photons.\ As another example of disentangling photons, we note that the spectral function $f_{4,C3}$ of Eq. (\ref{C3}) allows for the least entangled biphoton source.\ By annihilating a pair of photons of either $\hat a_{s'(i')}$ or $\hat a_{s''(i'')}$, we collapse $f_{4,C3}$ into two single Gaussian functions.\ Choosing $\Delta\omega_{s',s''}$ $=$ $0$ in $f_{4,C3}$, in Fig. \ref{fig6} we show the collapsed spectral distribution of two idler photons, which has extremely low entropy of entanglement $S$ $=$ $0.028$.\ The eigenvalues $\lambda_n$ are plotted in a logarithmic scale, showing an abrupt decrease of Schmidt numbers.\ The first two eigenvalues are $0.997$ and $0.0028$, occupying most of the modes (up to $99.98\%$) in this biphoton source.\ In Fig. \ref{fig6}(b), we show their mode probability densities.\ As expected the FWHM of the mode probability density follows the spectral distribution at the cut of $\Delta\omega_{i'(i'')}$ $=$ $0$.\ The first mode of the idler $\hat a_{i''}$ has a shortened linewidth than the one of $\hat a_{i'}$ due to a squared Lorentzian function of $[\Gamma_3^{ N}/2-i(\Delta\omega_{i''})]^{-2}$ in $f_{4,C3}$ with $\Delta\omega_{s''}$ $=$ $0$.

We can also manipulate $S$ by modifying the driving pulse durations $\tau_a$ and $\tau_b$.\ When we increase them to $\tau_a$ $=$ $\tau_b$ $=$ $1\Gamma_3^{-1}$ in the projected  $f_{4,C3}$ as in Fig. \ref{fig6}, we find $S$ becomes $0.13$, which allows for a more entangled source since the entangling Gaussian function $e^{-(\Delta\omega_{i'}+\Delta\omega_{i''})^2\tau_{eff}^2/8}$ has a tighter photon correlation on the distribution axis $\Delta\omega_{i'}$ $=$ $-\Delta\omega_{i''}$.\ On the other hand when we set $\tau_a$ $=$ $0.25\Gamma_3^{-1}$ and $\tau_b$ $=$ $1\Gamma_3^{-1}$ in the asymmetric setting, we find $S$ becomes $0.023$, which is even smaller than the symmetric case in Fig. \ref{fig6}.\ This reflects the competition of this entangling and two other disentangling Gaussian functions.\ Though $\tau_{eff}$ still increases in the asymmetric setting, a more confined distribution from $\tau_b$ limits the overall projected spectral function distribution, thus decreasing $S$.

\section{Discussion and conclusion}

The multiphoton states generated from the cascade emissions in the scheme of cascaded AE are advantageous in the perspective of well-controlled AE preparations.\ Large-scale implementation of such cascaded scheme is feasible when preparing the atoms either in free-space, multiplexed setting \cite{Lan2009, Jen2016a}, or optical lattices \cite{Wang2015}, but will suffer from the low generation rate.\ Low generation rate results from the weak excitations and few-photon level of emissions.\ Also low light is more subject to propagation attenuation, making the detection of photons difficult.\ However this can be overcome by raising the repetition rates in excitations.\ Finite spectral windows for collapsing the multiphoton spectral functions are not considered here.\ But we expect of no significant modifications in their spectral properties qualitatively, where essentially the finite spectral window of projection averages out the spectral distribution.\ A gate time of several hundreds of nanoseconds in the photon counting device would be enough to neglect the effect from finite detection windows.

The other merit of the multiphoton source in the cascaded scheme is flexibility to entangle the photons either in signal or idler frequencies, making a repertoire of many possible entangled multiphoton sources.\ The signal frequency is best for optical fiber transmission while the idler one is preferential in quantum storage.\ In our scheme a highly entangled photon source can be generated by increasing the excitation pulse durations in the symmetric setting or tighten the spectral distribution on the axis $\Delta\omega_s$ $+$ $\Delta\omega_i$ $=$ $0$ that conserves the biphoton energy.\ By appropriately projecting out the photon  counterparts in the multiphoton spectral function, the entropy of entanglement and the spectral mode functions can be modified and manipulated to serve different purposes requiring either pure or entangled states.

For more entangled biphoton source, we have demonstrated a larger entropy of entanglement $S$ $=$ $2.37$ or $1.79$ from the projected three-photon states compared to $1.33$ \cite{Jen2012-2} from just one AE under the same driving conditions.\ This shows that a more entangled biphoton source can be generated from a multiphoton source with even larger capacity in the genuine multipartite entanglement.\ Our cascaded scheme here can also combine with the multiplexed one \cite{Jen2016a}.\ The multiplexed scheme manipulates the spectral property of the biphoton state by modifying their central frequencies or phases \cite{Jen2016b}.\ Its maximal entropy of entanglement $S_{M}$ can be described by $S$ $+$ $S_d$.\ $S$ is the entropy of entanglement from one AE while $S_d$ $\equiv$ $\log_2 (N_{\rm MP})$ with $N_{\rm MP}$, the number of multiplexed AEs.\ Therefore the multiplexed scheme increases $s_d$ as $N_{\rm MP}$ increases.\ Meanwhile the cascaded scheme enables a multiphoton source from sequentially-coupled AEs using diamond configurations.\ Its bipartite entanglement can be extracted from the biphoton states collapsed from the multiphoton ones.\ This way the cascaded scheme modifies and manipulates $S$ effectively, making the combination of these two schemes a full control of $S_{M}$.\ In addition to the spectral shaping of the cascade emissions by modifying driving conditions \cite{Jen2012-2} or multiplexing AEs \cite{Jen2016a, Jen2016b}, the cascaded scheme here provides an alternative route to spectrally shape an even more entangled biphoton source, thus overcomes the limitation of $S$ in the multiplexed scheme.

In conclusion, we propose a cascaded scheme to generate a multiphoton source from the cascade emissions of the atomic ensembles.\ Highly spectrally entangled ($k+1$)-photon source can be created using $k$-photon state as the seed along with an appropriate driving field either in the lower or upper transition of the diamond configuration.\ Under the FWM condition, this highly directional and frequency-correlated photon source are useful for quantum information processing and applicable to multimode quantum communication.\ Furthermore such entangled multiphoton source can be spectrally shaped with controllable driving conditions and ensemble properties (for example atomic density and geometry), which could potentially be implemented in quantum spectroscopy \cite{Dorfman2016}.

\section*{ACKNOWLEDGMENTS}
This work is supported by the Ministry of Science and Technology (MOST), Taiwan, under Grant No. MOST-103-2112-M-001-011.

\appendix
\section{Other routes for four-photon state generation}
In Sec. III.B, we have investigated four-photon state generation in the cascaded scheme, and demonstrated the first route $C1$.\ Other four possible routes to generate four-photon states can use three-photon states $|\Psi\rangle_{3,B1}$ and $|\Psi\rangle_{3,B2}$ as seeds.\ The route $C2$ below is to annihilate $\hat a_{s'}^\dag$ of $|\Psi\rangle_{3,B1}$ with an extra coupling field $\Omega_a$ in the third AE'' to generate a newly correlated pair of photons $\hat a_{s''}^\dag \hat a_{i''}^\dag$.\ The effective state is $|\Psi\rangle_{4,C2}$ $=$ $f_{4,C2}\hat{a}^\dag_s\hat a_{i'}^\dag\hat a_{s''}^\dag\hat a_{i''}^\dag|0\rangle$ involving two signal and two idler photons with the spectral function
\begin{widetext}
\bea
\frac{f_{4,C2}}{\Gamma_3^3}=\frac{e^{-(\Delta\omega_s+\Delta\omega_{i'}+\Delta\omega_{s''}+\Delta\omega_{i''})^2\tau_{eff}^2/8}e^{-(\Delta\omega_{i'}+\Delta\omega_{s''}+\Delta\omega_{i''})^2\tau_b^2/4}e^{-(\Delta\omega_{s''}+\Delta\omega_{i''})^2\tau_a^2/4}}
{\big[\frac{\Gamma_3^{ N}}{2}-i(\Delta\omega_{i'}+\Delta\omega_{s''}+\Delta\omega_{i''})\big](\frac{\Gamma_3^{ N}}{2}-i\Delta\omega_{i'})(\frac{\Gamma_3^{ N}}{2}-i\Delta\omega_{i''})}.
\eea
The route $C3$ involves two signal and two idler photons, generated from a symmetric coupling between $\hat a_{s(i)}^\dag$ and pump fields $\Omega_{a(b)}$ in the third AE'' using three-photon states from routes $B1(2)$ respectively.\ Its spectral function has been shown in the main paper.\ For route $C4$, which annihilates $\hat a_{i'}^\dag$ of $|\Psi\rangle_{3,B2}$ with an extra coupling field $\Omega_b$ in the third AE''.\ The effective state is $|\Psi\rangle_{4,C4}$ $=$ $f_{4,C4}\hat{a}^\dag_i\hat a_{s'}^\dag\hat a_{s''}^\dag\hat a_{i''}^\dag|0\rangle$ involving two signal and two idler photons with the spectral function
\bea
\frac{f_{4,C4}}{\Gamma_3^3}=\frac{e^{-(\Delta\omega_{s'}+\Delta\omega_{i}+\Delta\omega_{s''}+\Delta\omega_{i''})^2\tau_{eff}^2/8}e^{-(\Delta\omega_{s'}+\Delta\omega_{s''}+\Delta\omega_{i''})^2\tau_a^2/4}e^{-(\Delta\omega_{s''}+\Delta\omega_{i''})^2\tau_b^2/4}}
{(\frac{\Gamma_3^{ N}}{2}-i\Delta\omega_{i})\big[\frac{\Gamma_3^{ N}}{2}-i(\Delta\omega_{s''}+\Delta\omega_{i''})\big](\frac{\Gamma_3^{ N}}{2}-i\Delta\omega_{i''})}.
\eea

The last route $C5$ annihilates $\hat a_{s'}^\dag$ of $|\Psi\rangle_{3,B2}$ with an extra coupling field $\Omega_a$ in the third AE''.\ The effective state is $|\Psi\rangle_{4,C5}$ $=$ $f_{4,C5}\hat{a}^\dag_i\hat a_{i'}^\dag\hat a_{s''}^\dag\hat a_{i''}^\dag|0\rangle$ involving one signal and three idler photons with the spectral function
\bea
\frac{f_{4,C5}}{\Gamma_3^3}=\frac{e^{-(\Delta\omega_{i}+\Delta\omega_{i'}+\Delta\omega_{s''}+\Delta\omega_{i''})^2\tau_{eff}^2/8}e^{-(\Delta\omega_{i'}+\Delta\omega_{s''}+\Delta\omega_{i''})^2\tau_a^2/4}e^{-(\Delta\omega_{s''}+\Delta\omega_{i''})^2\tau_a^2/4}}
{(\frac{\Gamma_3^{ N}}{2}-i\Delta\omega_{i})(\frac{\Gamma_3^{ N}}{2}-i\Delta\omega_{i'})(\frac{\Gamma_3^{ N}}{2}-i\Delta\omega_{i''})}.
\eea
\end{widetext}


\end{document}